\documentclass[11pt,twoside]{article}
\usepackage{asp2010}
\usepackage{graphicx}

\resetcounters

\begin{document}

\title{Measuring reliable surface rotation rates from \emph{Kepler} photometric observations}
\author{R.~A. Garc\'\i a$^1$, T. Ceillier$^1$, S. Mathur$^{2, 1}$, D. Salabert$^1$
\affil{$^1$La\-bo\-ra\-toi\-re AIM, CEA/DSM -- CNRS - Universit\'e Paris Diderot -- IR\-FU\-/SAp, 91191 Gif-sur-Yvette Cedex, France}
\affil{$^2$High Altitude Observatory, NCAR, P.O. Box 3000, Boulder, CO 80307, USA.}
}
\begin{abstract}
High-quality time series provided by space instrumentation such as CoRoT and {\it Kepler}, allow us to measure modulations in the light curves due to changes in the surface of stars related to rotation and activity. Therefore, we are able to infer the surface -- possibly differential -- rotation rate. However, instrumental perturbations can also produce artificial modulations in the light curves that would mimic those of truly stellar origin. In this work we will concentrate on {\it Kepler} observations in order to review an optimal way to extract reliable surface rotation rates. 

\end{abstract}

\section{Introduction}
The surface rotation rate of a star is a function of its mass and evolutionary state. Therefore, gyrochronology \citep[e.g.][]{2007ApJ...669.1167B,2011ApJ...733L...9M} takes advantage of these properties to use the stellar rotation rate as a clock, allowing to infer stellar ages. Moreover, stellar structures are modified by the mixing induced by rotation  \citep[e.g.][]{1997ARA&A..35..557P,2004A&A...425..229M,2005A&A...440..653M}. Taking into account such rotation rates in the evolutionary codes is challenging in stellar physics. Indeed, the rotation rates obtained by state-of-the-art modeling \citep[e.g.][]{2010ApJ...715.1539T,2012AN....333..971C,2012A&A...544L...4E,2013EPJWC..4303012D,2013A&A...549A..74M} are much higher than what seismic observations of the Sun \citep{ThoToo1996,GarCor2004,2007Sci...316.1591G,2008AN....329..476G,2008A&A...484..517M,2012SoPh..tmp..149E} and other stars are starting to show \citep{2012Natur.481...55B,2012ApJ...756...19D,2012A&A...548A..10M}.

When stars are magnetically active, spots could develop at their surfaces. Spots are cooler than the quiet star and therefore they produce a reduction of its luminosity. Therefore long stellar photometric observations --such as the ones required for asteroseismic analyses-- are modulated by the appearance of these spots at a period of the average stellar rotation at the active latitudes.
The analysis of the long CoRoT \citep{2006cosp...36.3749B} and {\it Kepler} \citep{2010Sci...327..977B} light curves can be used to unveil the surface -- possibly differential -- rotation rate when the observations are sufficiently long \citep[e.g.][]{2009A&A...506..245M,2009A&A...506...41G,2011ApJ...733...95M,2011A&A...530A..97B,2013A&A...549A..12M,2013A&A...550A..32M,2013arXiv1303.6787M}, as well as to deduce a proxy of the magnetic activity of the star \citep{2010Sci...329.1032G,2011ApJ...732L...5C,2013ApJ...769...37B}. 

In this work we review the different techniques used to extract the surface rotation from photometric measurements as well as the problems encountered when {\it Kepler} data are used.

\section{The importance of the \emph{Kepler} dataset used}

\begin{figure*}[!htb]
\centering
\includegraphics[scale=0.26]{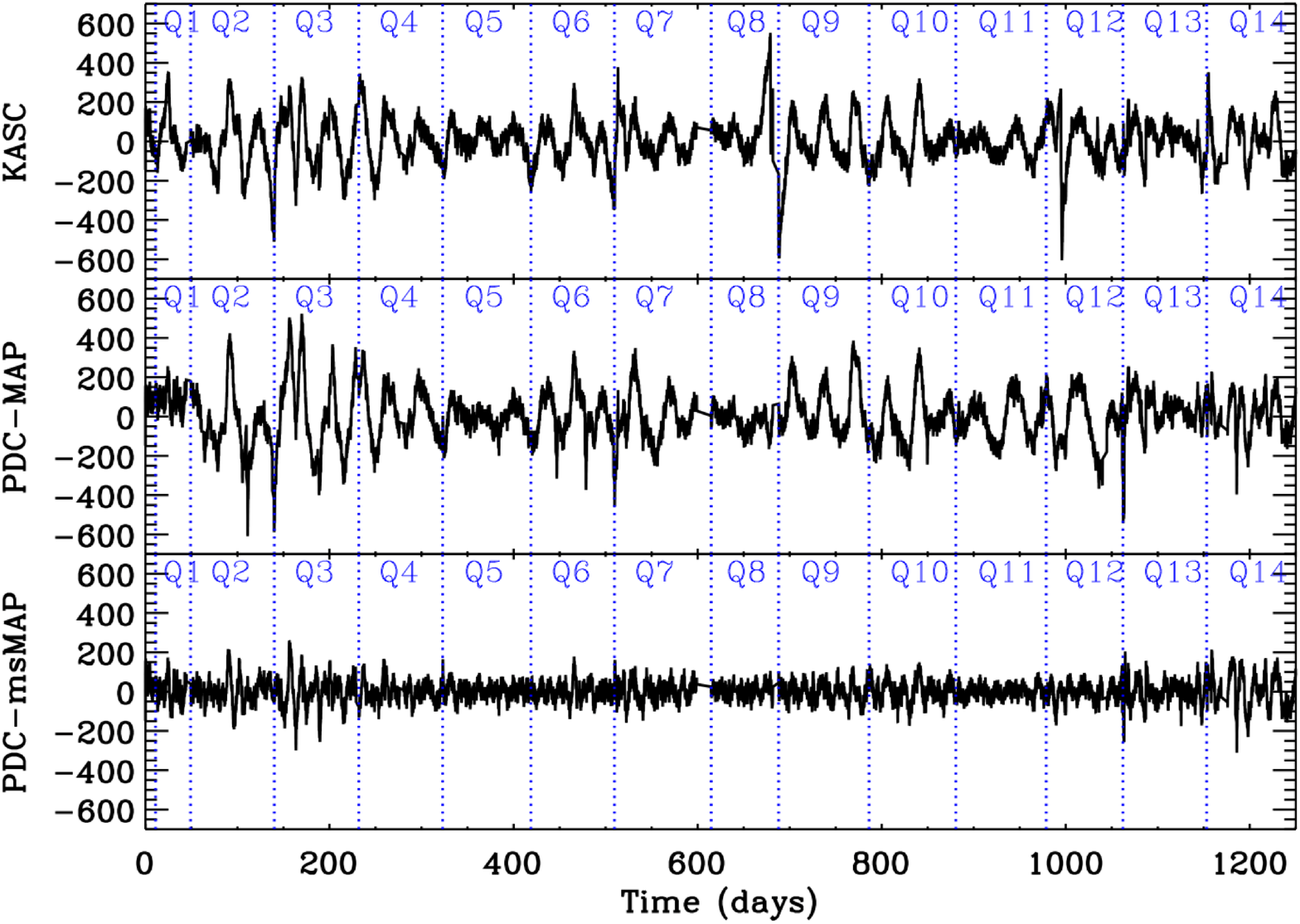}
\caption{Comparison of the light curves of KIC~11026764 computed using the three corrections of the {\it Kepler} data: PDC-MAP, PDC-msMAP and the KASC corrections. See the text for more details.}
\label{Fig0}
\end{figure*}

The {\it Kepler} mission is providing automated photometry for more than 150,000 stars at a time. This is not an easy task and the data analysis team is investing a huge amount of time to provide the best data products to satisfy many different scientific  communities, which have different requirements. Besides the Simple Aperture Photometry (SAP) time series, the project is now providing PDC (Pre-search Data Conditioning) processed data that are optimized for looking for transits in the light curves. The original Least-Squares PDC (PDC-LS) products, were changed by a Bayesian Maximum A Posteriori approach where a subset of highly correlated and quiet stars is used to generate a set of cotrending basis vector (CBV), which is then used to establish a range of ÒreasonableÓ robust fit parameters containing the common ``instrumental'' features \citep{2012arXiv1203.1382S,2012PASP..124.1000S}. An example is given in Fig.\ref{Fig0}.

 Recently, a major improvement of these datasets has been released: PDC-msMAP, in which ``msMAP'' stands for multi-scale MAP \citep{ThompsonRel21}. ``msMAP'' is a wavelet-based band-splitting framework for removing systematics from the light curves at different scales. Indeed, the light curves are first decomposed into three characteristics length-scales in which a standard PDC-MAP is independently applied. In the longest band ($>$ 21 days), the light curves are simply fitted to the CBVs  because it is assumed that the stellar signal cannot be distinguished from systematic instrumental effects, thus all the signal with a period larger than 21 days is removed.

A third correction system was developed by the KASC team following the algorithms used to correct the helioseismic data obtained by the GOLF/SoHO instrument \citep{GabGre1995,GarSTC2005}. This procedure \citep{2011MNRAS.414L...6G} allowed to correct for outliers and instrumental drifts as well as to stitch together the data from different quarters. An example of the residuals obtained with the three correction processes is shown in Fig.~\ref{Fig0} for the star KIC~11026764 \citep[an extensive analysis of this star can be found in][]{2010ApJ...723.1583M}. This star shows an average surface rotation rate at the active latitudes of 33 days.
Although in this particular example,  PDC-MAP and KASC corrections provide quite similar results, the effect of the high-pass filter is clearly seen in the PDC-msMAP light curve. The signature of the 33 days modulation has been removed but there is still a modulation of 12.5 days in the luminosity, which is an alias of the original rotation period. This must be taken into account when these data product are used to extract rotation periods.

Finally it is important to note that PDC-MAP and the KASC corrections could fail in some stars. Because of the differences in the procedures used, these failures would normally happen for different stars, and using both correction systems could improve the robustness of the extracted rotation periods, while minimizing the impact of the instrumental problems. 

\section{Extracting the surface rotation rates}
The techniques used to extract the stellar surface rotation rates can be categorized into three groups: a) spot modeling; b) studying the low-frequency part of the periodogram; c) using the autocorrelation of the temporal signal.

Spot modeling consists in parametrizing the evolution of spots by a given analytical model \citep[e.g.][]{1987ApJ...320..756D}, or  by integrating the stellar surface that has been divided into a pixelated grid \citep[e.g.][]{2010A&A...520A..53L}, and then fit it to the observations. Examples of some recent spot modeling fits to CoRoT and {\it Kepler} data can be found in \citet{2009A&A...506..245M,2012A&A...543A.146F} and \citet{2013ApJS..205...17W}. In general, we could say that when spot modelling is used, the extraction of the average rotation rate, as well as the spot lifetimes are robust, although the estimations of the sizes and distribution of spots, the inclination of the star and the differential rotation is more uncertain (B. Mosser private communication). The main problem of applying this technique to the actual flow of thousands of stars observed by CoRoT and {\it Kepler} is that it is very time consuming and it should be limited to a small fraction of stars.

A faster way to extract the surface rotation is by analyzing the low-frequency part of the periodogram and selecting the highest peak in a given frequency region \citep[e.g.][]{2009A&A...506...51B,2010A&A...518A..53M,2011A&A...534A...6C}, methodology that can be applied to thousands of stars \citep[e.g.][]{2013arXiv1305.5721N}. The main problem of using this method is that sometimes the highest peak in the periodogram is not the rotation period but the second or the third harmonic. An example of this phenomenon can be seen in the {\it Kepler} target KIC~4918333 as reported by  \citet{2013arXiv1303.6787M} (see Fig. 5 of the paper). In this case, the highest peak in the periodogram is the second harmonic. This could happen when there are two spots or active longitudes, one in the front and one in the back side of the star separated by approximately $180\deg$. As the region covered by the spots is not the same, the light curve shows half the period but alternating one peak larger than the following one. With this method it is important to check if there is another peak with enough power at double or triple the frequency of the highest peak. 

\begin{figure*}[!htb]
\centering
\includegraphics[scale=0.36]{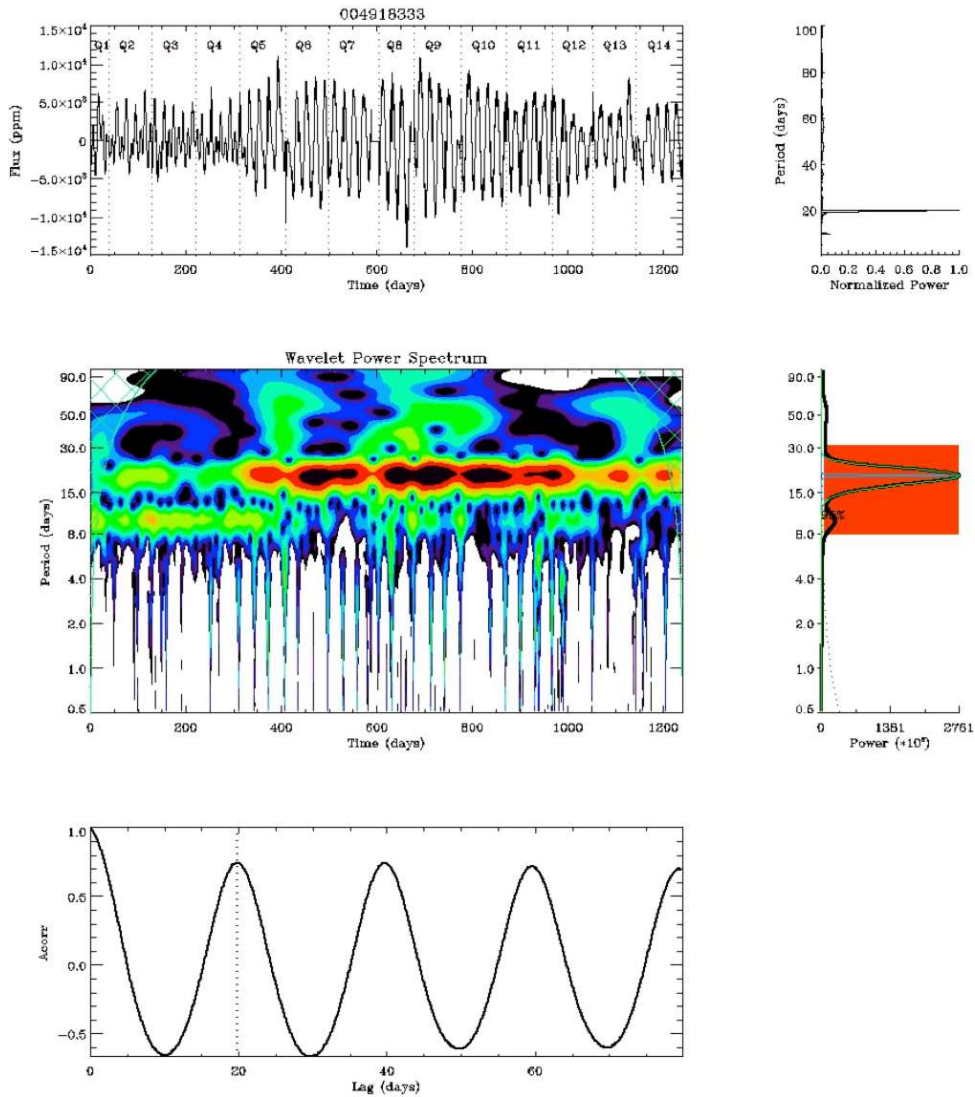}
\caption{Analysis of the {\it Kepler} star KIC~4918333 for quarters 0 to 14. In the top-left panel the smoothed light curved is shown. In the top-right panel, the Lomb-scargle periodogram of the full series is computed in a linear scale. The highest peak corresponds to the rotation period of the star: 20 days. The intermediate panel shows the time-frequency analysis using a wavelet power spectrum (left panel) as well as the projection into the period axis in logarithmic scale (right panel). The red shaded area corresponds to the region in which a Gaussian is fitted to extract the rotation period.}
\label{Fig1_McQ}
\end{figure*}

Another possibility is to use the autocorrelation of the temporal signal, or try to analyze longer time series. Indeed in the original analysis by \citet{2013arXiv1303.6787M}, only 4 {\it Kepler} quarters (1-4) were used. In that case, the autocorrelation of the temporal signal (after interpolating the data into a regular grid) provided the correct rotation period at 20 days, while the analysis of the periodogram gave the second harmonic, i.e., 10 days. However, when the analysis is done up to quarter 14, the ambiguity on the rotation period disappears as it is shown in Fig.~\ref{Fig1_McQ}. In the top-left panel, the PDC-MAP {\it Kepler} time series \citep{2012arXiv1203.1382S,2012PASP..124.1000S} of the first fifteen quarters are shown. The Fourier transform of the low-frequency part of the lomb-scargle periodogram is plotted in the top-right panel. The dominant peak is the one at 20 days and not the second harmonic at 10 days. 

\begin{figure*}[!htb]
\centering
\includegraphics[scale=0.36]{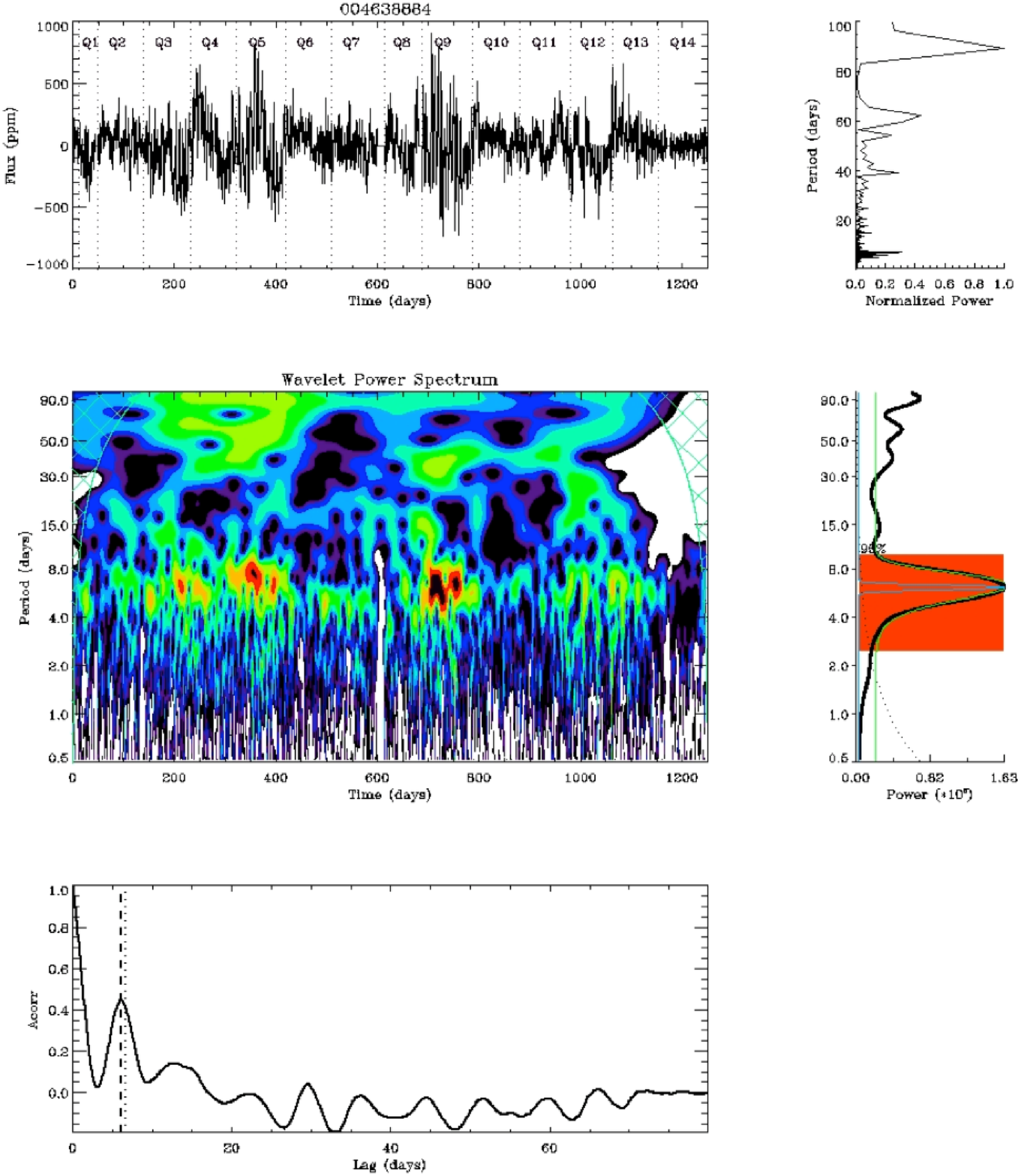}
\caption{Analysis of the {\it Kepler} star KIC~4638884 for quarters 0 to 14. Same legend than in Fig.~\ref{Fig1_McQ}.}
\label{Fig2}
\end{figure*}

The middle left panel in Fig.~\ref{Fig1_McQ} shows the time-frequency analysis using a wavelet transform based on the Morlet mother wavelet \citep{1998BAMS...79...61T}. Projecting the time-frequency diagram into the period domain we obtain the global wavelet power spectrum (GPWS, see intermediate-right panel of Fig.~\ref{Fig1_McQ}), where we can extract the rotation period by fitting a Gaussian function to the highest peak \citep{2010A&A...511A..46M}. With this methodology it is possible to verify if the periodicity is stable in time or if the main peak in the GWPS corresponds to a perturbation in the light curve. Indeed another problem of relying on the highest peak of the lomb-scargle periodogram --at low frequency-- is that it is very sensitive to instrumental problems in the data and sometimes the highest peak corresponds to a perturbation of the data (see Fig.~\ref{Fig2}). Having the time-frequency analysis helps to determine wether or not the retrieved period is very localized in time and corresponds to an instrumental perturbation. Moreover, the wavelet analysis takes into account the repetitivity of the signal and the highest peak is therefore the peak at 6 days and not the problem in the data at $\sim$ 80 days. 

Another way to avoid the perturbations in the light curve when using the periodogram is by doing the analysis into subseries of data \citep[e.g. by quarters as done by][]{2013arXiv1305.5721N}. However this implies to reduce the range of rotation periods that we are able to sample in a reliable way.


\section{Conclusion}
In this paper we have reviewed the main problems found to extract the correct rotation rates from the photometric measurements obtained by the {\it Kepler} mission. Indeed, from the 3 data products available, PDC-msMAP, PDC-MAP, and the KASC light curves, only the last two are suited for this kind of research. PDC-msMAP should be used with caution as it is stated in the associated documentation. Alias frequencies of periods longer than the cut-off limit of 21 days can be wrongly extracted.

We have also reviewed the main techniques usually used to extract the rotation rates, i.e., the autocorrelation of the temporal signal and the measurement of the highest peak in the periodogram. For the latter technique, it is always important to verify if there is a signal at  lower frequencies because in some cases, the highest peak will not be the first harmonic of the rotation period but the second or the third. To minimize the influence of the instrumental perturbations in the light curves we have demonstrated that techniques based on time-frequency analysis, for example using wavelets, would contribute to minimize these non-desirable effects.

\acknowledgements 
Funding for the {\it Kepler} Discovery mission is provided by NA\-SA's Science Mission Directorate. The authors wish to thank the entire {\it Kepler} team, without whom these results would not be possible. The CoRoT space mission has been developed and is operated by CNES, with contributions from Austria, Belgium, Brazil, ESA (RSSD and Science Program), Germany and Spain.  The authors thank the support of the French PNPS program as well as the CNES for the support of the CoRoT and GOLF/SoHO activities at the SAp, CEA-Saclay. RAG and SM acknowledge the support of the European Community's Seventh Framework Programme (FP7/2007-2013) under grant agreement no. 269194 (IRSES/\-ASK). SM acknowledges the support of the University of Tokyo. NCAR is supported by the National Science Foundation. \\

\bibliographystyle{asp2010} 
\bibliography{Fujihara_Rgarcia_v2PS}

\end{document}